\documentclass[reqno,12pt]{amsart}
\usepackage{geometry} 
\usepackage{amsmath,amsthm,amsfonts,amscd,amssymb,mathrsfs,graphicx}
\usepackage{a4wide}
\usepackage{microtype}
\usepackage{bbm}
\usepackage{slashed}
\usepackage{times}
\usepackage{hyperref}

\usepackage{tikz}
\usetikzlibrary{matrix,arrows,shapes}
\usepackage{color}

\newcommand{\ph}{\varphi}

\newcommand{\RRR}{{\mathfrak R}}

\newcommand{\HHH}{{\mathcal H}}
\newcommand{\R}{{\mathcal R}}

\newcommand{\NN}{{\mathcal N}}

\newcommand{\bm}[1]{\mathbf{#1}}
\def\1{\mathbbm{1}}\def\1{\mathbbm{1}}

\theoremstyle{definition}

\title{Cosmological perturbation theory and quantum gravity}
\author[R. Brunetti, K. Fredenhagen, T-P. Hack, N. Pinamonti, K. Rejzner]{Romeo Brunetti$^{1,a}$, Klaus Fredenhagen$^{2,b}$, Thomas-Paul Hack$^{3,c}$, Nicola Pinamonti$^{4,d}$ and Katarzyna Rejzner$^{5,e}$}
\date{} 

\begin{document}

\maketitle

\begin{center}
{\footnotesize 
$^1$Dipartimento di Matematica, Universit\`a di Trento, \\
$^2$II Institute   f\"ur Theoretische Physik, Universit\"at Hamburg,  \\
$^3$Institute f\"ur Theoretische Physik, Universit\"at Leipzig, \\
$^4$Dipartimento di Matematica, Universit\`a di Genova, and INFN, Sezione di Genova, \\
$^5$Department of Mathematics, University of York.
\par 
\rm\large\large 
\bigskip
\noindent\footnotesize 
Email: 
$^A$brunetti@science.unitn.it,
$^B$klaus.fredenhagen@desy.de,
$^C$thomas-paul.hack@itp.uni-leipzig.de,
$^D$pinamont@dima.unige.it,
$^E$kasia.rejzner@york.ac.uk
}
\end{center}

\begin{abstract}
It is shown how cosmological perturbation theory arises from a fully quantized perturbative theory of quantum gravity. Central for the derivation is a non-perturbative concept of gauge-invariant local observables by means of which perturbative invariant expressions of arbitrary order are generated. In particular, in the linearised theory, first order gauge-invariant observables familiar from cosmological perturbation theory are recovered. Explicit expressions of second order quantities are presented as well. 
\end{abstract}
\section{Introduction}
The fluctuations of the cosmological microwave background provide a deep insight into the early history of the universe. The most successful theoretical explanation is inflationary cosmology where a scalar field (the inflaton) is coupled to the gravitational field. Usually, the theory is considered in linear order around a highly symmetric background, typically the spatially flat Friedmann-Lema\^ itre-Robertson-Walker spacetime. 

Extending the theory to higher orders is accompanied by severe obstacles. Already in a classical analysis the definition of gauge-invariant observables turns out to be rather complicated; moreover, one is immediately confronted with the problem of constructing a theory of quantum gravity. Previous treatments of higher-order cosmological perturbation theory include \cite{Bartolo:2006fj, Bruni:1996im, Langlois:2010vx, Maldacena:2002vr, Malik:2008im, Nakamura:2004rm, Noh:2004bc, Hwang:2012aa}; many further references on the subject can be found e.g. in \cite{Langlois:2010vx}.

In a recent paper \cite{BFR} three of us reanalysed the field theoretical construction of quantum gravity from the view point of locally covariant quantum field theory. This analysis was based on the methods of perturbative Algebraic Quantum Field Theory (pAQFT), see \cite{FR-Advances in AQFT} and references therein, and on an adapted version of the Batalin-Vilkovisky formalism for the treatment of local gauge symmetries \cite{Hollands,FR-BVQ}. The result was that a consistent theory (in the sense of an expansion into a formal power series) 
exists and is independent of the background. Due to non-renormalisability, however, in each order of perturbation theory new dimensionful coupling constants occur, which have to be fixed by experiments; hence the theory should be interpreted as an effective theory that is valid at scales where these new constants are irrelevant.
One might hope that non-perturbative effects improve the situation in the sense of Weinberg's concept of asymptotic safety, since there are encouraging results supporting this perspective; see for example \cite{Reuter, Reuter2}. Furthermore, it is difficult to observe any effects of quantum gravity, so it seems reasonable to start from the hypothesis that at presently accessible scales the influence of these higher order contributions is small. 

One of the main questions addressed by \cite{BFR} in the construction of the theory was the existence of local observables. It was answered, in a way familiar from classical general relativity, by using physical scalar fields, e.g. curvature scalars, as coordinates, and by expressing other fields as functions of these coordinates. Since quantization in the framework
of pAQFT relies on a field theoretical version of deformation quantization of classical theories (first introduced in \cite{DF}), the classical construction can be transferred to the quantum realm.

The procedure works as follows. One selects $4$ scalar fields $X_{\Gamma}^a, a=1,\ldots 4$, which are functionals of the field configuration $\Gamma$ which includes the spacetime metric $g$, the inflaton field $\phi$ and possibly other fields. The fields $X_{\Gamma}^a$ are supposed to transform under diffeomorphisms $\chi$ as           
\begin{equation}\label{equivariance}
X_{\chi^*\Gamma}^a=X_{\Gamma}^a\circ\chi\ ,
\end{equation}
where $\chi^*$ denotes the pullback (of sections of direct sums of tensor products of the cotangent bundle) via $\chi$.
We choose a background $\Gamma_0$ such that the map
\begin{equation}
X_{\Gamma_0}:x\mapsto (X_{\Gamma_0}^1,\ldots,X_{\Gamma_0}^4)
\end{equation}
is injective. In order to achieve injectivity on cosmological backgrounds $\Gamma_0$, we shall be forced to include the coordinates $x$ in the construction of $X_\Gamma$ in a way which is compatible with  \eqref{equivariance}. We then consider $\Gamma$ sufficiently near $\Gamma_0$ and set
\begin{equation}
\alpha_{\Gamma}=X_{\Gamma}^{-1}\circ X_{\Gamma_0}\,.
\end{equation}
We observe that $\alpha_{\Gamma}$ transforms under diffeomorphisms -- which leave the background $\Gamma_0$, that is by definition fixed, invariant -- as
\begin{equation}
\alpha_{\chi^*\Gamma}=\chi^{-1}\circ\alpha_{\Gamma}\,.
\end{equation}
Let now $A_{\Gamma}$ be any other scalar field which is a local functional of $\Gamma$ and transforms under diffeomorphisms as in \eqref{equivariance}. Then the field
\begin{equation}
\mathcal {A}_{\Gamma}:=A_{\Gamma}\circ\alpha_{\Gamma}
\end{equation}
is invariant under diffeomorphisms and may be considered as a local observable. Note that invariance is obtained by shifting the argument of the field in a way which depends on the configuration. 

The physical interpretation of this construction is as follows: the fields $X^a_\Gamma$ are configuration-dependent coordinates such that $[A_{\Gamma}\circ X_{\Gamma}^{-1}](Y)$ corresponds to the value of the quantity $A_{\Gamma}$ provided that the quantity $X_{\Gamma}$ has the value $X_{\Gamma}=Y$. Thus $A_\Gamma \circ X_\Gamma^{-1}$ is a partial or relational observable \cite{Rovelli:2001bz, Dittrich:2005kc, Thiemann:2004wk}, and by considering $\mathcal {A}_{\Gamma} = A_\Gamma \circ X_\Gamma^{-1} \circ X_{\Gamma_0}$ we can interpret this observable as a field on the background spacetime.

Clearly, to make things precise, one also has to characterise the region in the configuration space where all the maps are well defined and restrict oneself to configurations $\Gamma$ in the appropriate neighbourhood of the background $\Gamma_0$, see \cite{BFR,Igor} for details.

Fortunately, in formal deformation quantization as well as in perturbation theory, only the Taylor expansion of observables around some background configuration enters, hence it is sufficient to establish the injectivity of $X_{\Gamma_0}$ in order for the expansion of $\mathcal{A}_{\Gamma_0+\delta\Gamma}$ around $\Gamma_0$ to be well-defined.
As an example we compute this expansion up to the first order. We obtain 
\begin{equation}
\mathcal{A}_{\Gamma_0+\delta\Gamma}=A_{\Gamma_0}+\left\langle\frac{\delta A_{\Gamma}}{\delta \Gamma}(\Gamma_0),\delta\Gamma\right\rangle+\frac{\partial A_{\Gamma_0}}{\partial x^{\mu}}\left\langle\frac{\delta\alpha_{\Gamma}^{\mu}}{\delta\Gamma}(\Gamma_0),\delta\Gamma\right\rangle + O(\delta \Gamma^2)\ .
\end{equation}
The third term on the right hand side is necessary in order to get gauge-invariant fields (up to first order). 
We calculate
\begin{equation}
\frac{\delta\alpha_{\Gamma}^{\mu}}{\delta\Gamma}(\Gamma_0)=-\left(\left(\frac{\partial X_{\Gamma_0}}{\partial x}\right)^{-1}\right)^{\mu}_ a\frac{\delta X_{\Gamma}^a}{\delta\Gamma}(\Gamma_0)\ .
\end{equation}

In this work we apply this general idea to inflationary cosmology. In contrast to other systematic or covariant attempts to define gauge-invariant quantities in higher-order cosmological perturbation theory, see for example \cite{Langlois:2010vx, Malik:2008im, Nakamura:2014kza, Hwang:2012aa}, our construction works off-shell, is based on a clear and simple concept which is applicable to general backgrounds such that cosmological perturbation theory may be viewed as a particular application of perturbative quantum gravity \cite{BFR}. Moreover, we construct non-perturbative gauge-invariant quantities whose perturbative expansion to arbitrary orders may be computed algorithmically without the need for additional input at each order.

This paper is organised as follows: In the second section we recall a few  basic facts about perturbation theory of the Einstein-Klein-Gordon system on cosmological backgrounds. In the third section we describe the general method to obtain gauge invariant observables at all orders on generic backgrounds. We furthermore discuss how to treat the case of a FLRW background where the large symmetry prevents us from using coordinates constructed from the dynamical fields alone. The fourth section contains the analysis of two gauge invariant observables at second order. The steps necessary for the construction of a full all-order quantum theory are  briefly sketched in Section 5. Finally a number of conclusions are drawn in the last section.

 \section{Perturbations of the Einstein-Klein-Gordon system on a FLRW spacetime}
 \label{sec:perturbationsintro}

We consider the Einstein-Klein-Gordon system, namely a minimally coupled scalar field $\tilde\phi$ with potential $V(\tilde\phi)$ propagating on a Lorentzian spacetime $(M,\tilde g)$ with field equations
\begin{equation}\label{eq:EKG}
R_{ab}-\frac{1}{2}R \tilde g_{ab} =  T_{ab}    ,\qquad     - \Box \tilde\phi + V^{(1)}(\tilde\phi) = 0,
\end{equation}
where $T_{ab}$ is the stress tensor of $\tilde\phi$, $R_{ab}$ the Ricci tensor and $R$ the Ricci scalar. 
We discuss perturbations of this system around a background. 
A linearised theory is obtained starting from a one-parameter family of solutions $\lambda \mapsto \Gamma_\lambda :=  (\tilde g_{\lambda},\tilde\phi_{\lambda})$ and considering
\[
\delta \Gamma := (\gamma,\varphi) := \left. \frac{d}{d\lambda}(\tilde g_{\lambda},\tilde\phi_{\lambda}) \right|_{\lambda = 0},
\] 
hence $\Gamma_0:=( g,\phi):=(\tilde g_0,\tilde\phi_0)$ is the background configuration while $\delta \Gamma = (\gamma,\varphi)$ is the linearised perturbation.

The background solution we choose consists of a flat Friedmann-Lema\^ itre-Robertson-Walker (FLRW) spacetime $(M,g)$
together with a scalar field $\phi$ which is constant in space.
We recall that a flat FLRW spacetime is conformally flat and that 
\begin{equation}\label{eq:FRW}
M = I \times \mathbb{R}^3, \qquad g= a^2(\tau)(-d\tau\otimes d\tau+\sum_i d{x}^i\otimes d{x}^i ),
\end{equation}
where $I\subset \mathbb{R}$ is an open interval, the scale factor $a(\tau)$ is a function of the conformal time $\tau$ and where $x^i$ are three-dimensional Cartesian (comoving) coordinates.
The background equations of motion of the system are best displayed in terms of the auxiliary function 
\[
\mathcal{H} := \frac{a'}{a} ,   
\]
where $a'$ indicates the derivative with respect to the conformal time. $\mathcal{H}$ is related to the Hubble parameter 
$H=\mathcal{H}a^{-1}$ and to the Ricci scalar $R = 6(\mathcal{H}'+ \mathcal{H}^2) a^{-2}$.
The background equations of motion are
\begin{gather*} 
\mathcal{H}^2= (\phi')^2+2a^2 V(\phi),  \qquad
2(\mathcal{H}'+ 2\mathcal{H}^2) = - (\phi')^2+2a^2 V(\phi),\\
\phi'' + 2\mathcal{H} \phi'+ a^2V^{(1)}(\phi) = 0.
\end{gather*}

A generic perturbation $\gamma$ of the FLRW metric $g$ can be decomposed in the following way
\begin{equation}\label{eq:perturbation}
\gamma = 
a(\tau)^2\begin{pmatrix}
-2A &&  (-\partial_i B   + V_i)^t \\
-\partial_i B   + V_i  &&   2(\partial_i\partial_j E+\delta_{ij}D+\partial_{(i}W_{j)}+T_{ij})
 \end{pmatrix}
\end{equation}
where $A,B,D,E$ are scalars, $V, W$ are three dimensional vectors and $T$ is a tensor on 3-dimensional Euclidean space.  
The decomposition is unique if all these perturbations vanish at infinity  and if
\[
{T_i}^i = 0 , \qquad \partial_i {T^i}_j = 0 , \qquad \partial_i {V^i} = 0 , \qquad \partial_i {W^i} = 0
\]
(see e.g. Proposition 3.1 in \cite{Hack}).

Under an infinitesimal first order gauge transformation the linear perturbations transform in the following way
\[
\gamma_{ab} \mapsto  \gamma_{ab} + \mathcal{L}_{\xi} g_{ab} = \gamma_{ab} + 2\nabla_{(a} \xi_{b)} , \qquad  
\varphi \mapsto  \varphi + \mathcal{L}_{\xi} \phi = \varphi + \xi(\phi).  
\]
In particular 
\begin{gather*}
A\mapsto A+(\partial_\tau+\mathcal{H})r, \qquad 
B\mapsto B+r-s', \qquad 
D\mapsto D+\mathcal{H}r, \qquad 
E\mapsto E+s, \\
\varphi\mapsto \varphi+\phi'r, \qquad
V_i\mapsto V_i + v_i', \qquad 
W_i\mapsto W_i + v_i, \qquad 
T_{ij}\mapsto T_{ij}, \qquad 
\end{gather*}
where the generator $\xi$ of one-parameter gauge transformations is also decomposed as
\begin{equation}\label{eq:perturbation_diffeo}
\xi^0 = r,\qquad \xi^i = \partial_i s + v_i, \qquad   \partial_i v^i = 0.
\end{equation}
Notice that the gauge transformations do not mix scalar, vector or tensor perturbations at linear order.

Furthermore, we observe that tensor perturbations are gauge-invariant and that gauge-invariant vector perturbations can be obtained considering $X_i := W'_i-V_i$. Regarding the scalar perturbations we see that the following fields are gauge-invariant
\begin{equation}\label{eq:Bardeen}
\Phi:= A-(\partial_t+\mathcal{H})(B+E'), \qquad
\Psi:= D- \mathcal{H}(B+E'), \qquad \chi := \varphi - \phi' (B+E').
\end{equation}
The first two of them are called Bardeen potentials.

Let us recall the form of the linearised equations of motions satisfied by the gauge-invariant perturbations. 
The first observation is that the equations of motion respect the decomposition in scalar, vector and tensor perturbations. 
In particular, for the vector and tensor perturbations, it holds that 
\begin{equation}\label{eq:linear-equations}
\Delta X_i =0, \qquad   (\partial_t + 2\mathcal{H})X_i =0, \qquad
\frac{1}{a^2} (\partial_t^2 +2 \mathcal{H} \partial_t - \Delta ) T_{ij}  =0.
\end{equation}
For the scalar part the equations of motion are better displayed in terms of the Mukhanov-Sasaki variable 
\begin{equation}\label{eq:muk-sas}
\mu := \chi  - \frac{\phi'}{\mathcal{H}} \Psi   = \varphi  - \frac{\phi'}{\mathcal{H}} D. 
\end{equation}
The equation of motion for this variable is decoupled also from the other scalars of the theory, in fact
\[
\left( -\Box +\frac{R}{6} - \frac{z''}{z a^2} \right) \mu = 0 , \qquad    z := \frac{a \phi'}{\mathcal{H}}.
\]
The other scalar perturbations can be obtained in terms of $\mu$.
In particular the Bardeen potential $\Phi$ is the unique solution of
\begin{equation}\label{eq:onshellfirstorder}
\Delta \Phi = \frac{\phi^\prime}{2}\left(\mu^\prime + \left(\frac{\mathcal{H}^\prime}{\mathcal{H}}-\frac{\phi^{\prime\prime}}{\phi^\prime}\right)\mu\right)
\end{equation}
while the other scalar perturbations are given by 
\begin{equation}\label{eq:onshellfirstorder2}
\Psi = -\Phi , \qquad
\chi = \frac{2}{\phi'} (\partial_\tau + \mathcal{H}) \Phi.  
\end{equation}

We briefly discuss the situation beyond linear order. According to \cite{Sonego:1997np}, infinitesimal diffeomorphisms may be approximated by so-called knight diffeomorphisms, which are of the form $\exp \mathcal{L}_{\xi}$ with $\xi = \lambda \xi_1 + \frac12 \lambda^2 \xi_2 + O(\lambda^3)$. Analogously we may expand a configuration $\Gamma$ as $ \Gamma = \Gamma_0+\delta \Gamma=\Gamma_0 + \lambda \delta\Gamma_1 + \frac12 \lambda^2 \delta\Gamma_2 + O(\lambda^3)$, and determine the transformation behaviour of separate orders by considering $\exp \mathcal{L}_{\xi} \, \Gamma$ at fixed order in $\lambda$, see for example \cite{Bartolo:2006fj, Bruni:1996im, Malik:2008im, Nakamura:2004rm, Noh:2004bc}. Assuming that $\xi$ and $\delta \Gamma$ vanish at spatial infinity, each order $\xi_i$ and $\delta \Gamma_i$ may be uniquely decomposed as in \eqref{eq:perturbation_diffeo} and \eqref{eq:perturbation}. The transformation behaviour of the components of the latter decomposition becomes more complicated than at linear order, since higher-order gauge transformations mix scalar, vector and tensor quantities in a non-local fashion, as do the higher-order equations of motion. We shall not be concerned with the explicit form of higher-order gauge transformations in this work, as our constructions do not rely on these details and the quantities we consider are manifestly all-order gauge-invariant from the outset.

For the remainder of this work we shall use the following notation motivated by the fact that the space of configurations is an affine space. We decompose a general configuration $\Gamma$ as $\Gamma := (\tilde g, \tilde \phi) :=\Gamma_0 + \delta \Gamma $, where $\tilde g := g + \gamma$, $\tilde \phi := \phi + \varphi$ and $\delta \Gamma := (\gamma,\varphi)$ effectively subsumes linear and higher orders of the perturbation of the background $\Gamma_0 := (g,\phi)$. This applies analogously to the components of the decomposition \eqref{eq:perturbation} of $\gamma$.

For later use we recall a useful observation regarding Bardeen potentials. The linear Bardeen potentials $\Phi$, $\Psi$ and the gauge-invariant scalar field perturbation $\chi$ in \eqref{eq:Bardeen} have the advantage that they coincide with $A$, $D$, and $\varphi$ respectively in the so-called longitudinal or conformal gauge where the components $B$ and $E$ of the metric perturbation $\gamma$ vanish. This gauge and the definition of the gauge-invariant quantities $\Phi$, $\Psi$ and $\chi$ may be extended to higher orders, such that also at higher orders $\Phi=A$, $\Psi = D$, $\chi=\varphi$ if $B=E=0$, see for example \cite{Malik:2008im}.

\section{All-order gauge-invariant observables on FLRW backgrounds}
\label{sec_covcoords}

In this section we provide details on the general construction of all-order gauge-invariant quantities on general and FLRW backgrounds before discussing examples in the next section.

In perturbative Algebraic Quantum Field Theory (pAQFT) -- the conceptual framework underlying perturbative quantum gravity in \cite{BFR} --  observables of a field theory are described as functionals of smooth field configurations $\Gamma=(\tilde g,\tilde\phi)$. For the purpose of cosmological perturbation theory, we need the additional restriction that configurations vanish at spatial infinity. In order to be able to operate on the functionals, some regularity is required: the functional derivatives to all orders should exist as distributions of compact support. 

Moreover, we restrict our attention to local functionals, i.e. those functionals whose $n-$th order functional derivatives are supported on the diagonal of $M^n$ for every $n$.
Examples of objects of this form are 
\begin{equation}\label{eq:field}
A_{\Gamma}(f):= \int_M A_{\Gamma} f 
\end{equation}
where $A_{\Gamma}$ is a smooth scalar function which is a polynomial in the derivatives of the field configuration 
$\Gamma=(\tilde g,\tilde\phi)$ (i.e. $A_{\Gamma}(x)=F(j_x(\Gamma))$ with $F$ a smooth function on the appropriate jet bundle) and where 
$f$ is a smooth compactly supported test density. However, later on in this work we are forced to consider also functionals which violate this locality condition as well as the condition of compact support. The diffeomorphisms $\chi$ of the spacetime act on configurations via pullback $\Gamma\mapsto\chi^*\Gamma$, 
and candidates for gauge-invariant fields are equivariant in the sense that
\begin{equation}\label{eq_equivariant}
A_{\chi^*\Gamma}=A_{\Gamma}\circ\chi\ .
\end{equation}
Thus in order to exhibit gauge-invariant functionals one has to consider test densities $f_{\Gamma}$ which depend on the field configuration $\Gamma$ such that
\begin{equation}
f_{\chi^*\Gamma}=\chi_*f_{\Gamma}\,,
\end{equation}
where $\chi_*$ is the pushforward of test densities via $\chi$.

As described in the Introduction, in the general case we solve the problem by choosing four scalar fields $X_{\Gamma}^a$ which constitute a coordinate system
$X_{\Gamma}$ for a given background $\Gamma_0$, and define the $\Gamma$-dependent diffeomorphism
\begin{equation}
\alpha_{\Gamma}=X_{\Gamma}^{-1}\circ X_{\Gamma_0}\ .
\end{equation}
For arbitrary test densities $f$, we may now consider the $\Gamma$-dependent test densities $f_{\Gamma}=\alpha_{\Gamma}{}_*f$ in order to obtain gauge-invariant observables $A_\Gamma(f_\Gamma)$ by means of \eqref{eq:field}. Equivalently, we may directly 
consider the gauge-invariant field
\begin{equation}\label{eq:ref3}
\mathcal{A}_{\Gamma}=A_{\Gamma}\circ\alpha_{\Gamma}\ .
\end{equation}

Scalars that can be used as coordinates on generic backgrounds $\Gamma_0$ are e.g.  traces  of powers of the Ricci operator $\bm{R}$
\begin{equation}\label{eq:ricci-scalars}
X_\Gamma^a:=\text{Tr} (\bm{R}^{a}),  \qquad a\in\{1,2,3,4\}
\end{equation}
(the operator which maps one forms to one forms and whose components are given in terms of the Ricci tensor ${R_{a}}^b$). 
 When other (matter) fields are present in the considered model, also these can serve as coordinates, 
e.g.,  in the case of a Einstein-Klein-Gordon system, the scalar field $\tilde \phi$. 

In view of renormalisation it is advisable to use coordinates $X_\Gamma$ which are local functionals of the configuration $\Gamma$. As we shall discuss in the following, this does not seem to be possible in cosmological perturbation theory on account of the symmetries of FLRW backgrounds $\Gamma_0$.

\subsection{Perturbative expansion up to second order}
\label{sec:generalexpansion2}

To illustrate the general procedure we compute the second order expansion of the gauge-invariant field $\mathcal{A}_\Gamma$ which was to first order described in the Introduction. 

We observe that we have to calculate the functional derivatives of the diffeomorphisms ${\alpha_{\Gamma}}$ with respect to $\Gamma$. 
We use the notation
\begin{equation}
\left\langle\frac{\delta^n}{\delta \Gamma^n}X_{\Gamma}(\Gamma_0),\delta\Gamma^{\otimes n}\right\rangle=:X_n\ ,\quad \left\langle\frac{\delta^n}{\delta\Gamma^n}\alpha_{\Gamma}(\Gamma_0),\delta\Gamma^{\otimes n}\right\rangle=:x_n
\end{equation}
and find up to second order
\begin{equation}
x_0^{\mu}(x)=x^{\mu}\,,\quad x_1^{\mu}=-J^{\mu}_aX^a_1\,,
\end{equation}
where $J$ is the inverse of the Jacobian of $X_{\Gamma_0}$, and
\begin{equation}
x_2^{\mu}=-J^{\mu}_aX_2^a-J^{\mu}_aJ^{\nu}_bJ^{\rho}_c\frac{\partial^2 X_0^a}{\partial x^{\nu}\partial x^{\rho}}X_1^bX_1^c+2J^{\mu}_aJ^{\nu}_b\frac{\partial X_1^a}{\partial x^\nu}X_1^{b}\ .
\end{equation}
We use an analogous notation for the Taylor expansions of the fields $A_{\Gamma}$ and $\mathcal{A}_{\Gamma}$ and find
\begin{equation}\label{eq_gaugeinvexp1}
\mathcal{A}_0=A_0\,,\quad \mathcal{A}_1=A_1+\frac{\partial A_0}{\partial x^{\mu}}x_1^{\mu}\,,
\end{equation}
and
\begin{equation}\label{eq_gaugeinvexp2}
\mathcal{A}_2=A_2+2\frac{\partial A_1}{\partial x^{\mu}}x_1^{\mu}+\frac{\partial A_0}{\partial x^{\mu}}x_2^{\mu}+\frac{\partial^2 A_0}{\partial x^{\mu}\partial x^{\nu}}x_1^{\mu}x_1^{\nu}\ .
\end{equation}
 
\subsection{Non-degenerate covariant coordinates on FLRW backgrounds}
\label{sec:coordinates}

In order to obtain these expansions we need a $4$-tuple of equivariant fields which define a non-degenerate coordinate system on the background $\Gamma_0$. This is possible in the generic case, e.g. by using the ansatz \eqref{eq:ricci-scalars}, but creates problems, if the background metric possesses non-trivial symmetries. This applies to the case of FLRW backgrounds $\Gamma_0$ where only time functions can be constructed out of the background metric $g$ and the background scalar field $\phi$. In the following we present a construction of non-degenerate coordinates which solves the above-mentioned problem at the expense of being non-local, albeit in a controlled way. Note that introducing additional external fields as reference coordinates like in the Brown-Kucha\v{r} model \cite{Brown:1994py} is not useful in the context of cosmological perturbation theory because these fields would appear in the final gauge-invariant expressions and thus an interpretation of these in terms of only the fundamental dynamical fields is difficult. The construction we present in the following does involve the comoving spatial coordinates $x^i$ of the FLRW spacetime as an external input. However the explicit dependence on $x^i$ disappears from the final expressions because these depend on $X_{\Gamma_0}$ only via its Jacobian.

The simplest choice of the time coordinate is provided by the inflaton field itself, so we set 
\begin{equation}\label{eq_coord0}
X_\Gamma^0 = \tilde \phi = \phi + \varphi\,.
\end{equation}
The construction of the spatial coordinates $X^i_\Gamma$ needs a bit of preparation. To this end, we consider the unit time-like vector
\begin{equation}\label{eq_nphi}
n_\phi = \frac{\tilde g^{-1}(d\tilde \phi,\cdot)}{\sqrt{|\tilde g^{-1}(d\tilde \phi,d\tilde \phi)}|} = \frac{1}{a}(1-A)\partial_\tau + \frac{1}{a}\left(\partial^i B -\frac{\partial^i \varphi}{\phi^\prime}\right) \partial_i + O(\delta \Gamma^2)
\end{equation}
and the tensor
\begin{equation}\label{eq_hphi}
h_\phi = \tilde g + \tilde g(n_\phi,\cdot) \otimes \tilde g(n_\phi,\cdot)\,,
\end{equation}
where $\partial^i := \partial_i := \partial/\partial x^i$ and $x^i$ for $i\in\{1,2,3\}$ are comoving spatial coordinates on the FLRW spacetime $(M,g)$. $n_\phi$ is a unit normal on the hypersurfaces of constant $\tilde \phi$ and $h_\phi$ is the induced metric on these hypersurfaces.

Let $\Delta_\phi$ denote the Laplacian for $h_\phi$ and $G_\phi$ its inverse, which we choose by imposing the boundary condition that the background value of $G_\phi$ is the Coulomb potential $G_\Delta$ with suitable factors of the scale factor $a$. We define and compute  
$$
\Delta_\phi := \Delta_0 + \delta\Delta\,,\qquad \Delta_0 := \frac{\Delta}{a^2}\,,\qquad \Delta := \sum^3_{i=1} \partial^2_i
$$
$$
\delta\Delta = -\lambda\left( \frac{ 2( D + \Delta E)\Delta 
   -(\partial^i ( D - \Delta E))\partial_i}{a^2}+\frac{(\Delta \varphi ) \partial_\tau+(\partial^i \varphi)(2\partial_\tau + \HHH) \partial_i}{a^2 \phi^\prime}\right)+O(\delta \Gamma^2)
$$
$$
G_\phi  := G_0 + \delta G\,,\qquad G_0 := a^2 G_\Delta\,, \qquad G_\Delta \circ \Delta = \1\quad \text{on functions that vanish at spatial infinity}\,,
$$
$$
\delta G = \sum^\infty_{n=1} (-1)^n G_0 \circ (\delta \Delta \circ G_0)^{\circ n} = - G_0 \circ \delta \Delta \circ G_0 + O(\delta\Gamma^2)\,.
$$
Using these objects, we obtain
\begin{equation}\label{eq:Ycoords}
Y_\Gamma^i := \left(1-G_\phi  \circ \Delta_\phi \right)x^i = x^i +  \partial_i ( E+G_\Delta\mathfrak{R})+O(\delta \Gamma ^2)\,,\qquad \mathfrak{R} := \frac{\HHH}{\phi^\prime}\mu\,.
\end{equation}

We observe that $Y_\Gamma^i$ are harmonic coordinates for $\Delta_\phi $ that we have constructed by means of $x^i$, i.e. harmonic coordinates for $\Delta_0$. The construction of $Y_\Gamma^i$ makes sense for all configurations $\Gamma$ which vanish at spatial infinity, but not in general. The restriction to this set of configurations from the outset is natural in the context of cosmological perturbation theory -- recall that the decomposition \eqref{eq:Bardeen} is unique only in this case -- and does not create problems for the pAQFT framework. For consistency, we have to restrict the class of infinitesimal diffeomorphisms we consider in the same manner. In fact, a straightforward computation reveals that the functionals $Y_\Gamma^i$ are equivariant with respect to all diffeomorphisms $\chi$ that vanish at spatial infinity
$$
\chi^* Y_\Gamma^i = Y_{\chi^*\Gamma}^i + (1-G_{\chi^* \phi}\circ \Delta_{\chi^* \phi})(\chi^* x^i - x^i ) =  Y_{\chi^*\Gamma}^i\,,
$$
but not with respect to arbitrary diffeomorphisms. Here $\Delta_{\chi^* \phi}$ denotes the Laplacian constructed analogous to $\Delta_{\phi}$ but with $\chi^* \tilde \phi$ instead of $\tilde \phi$ and $G_{\chi^* \phi}$ denotes its inverse with the discussed boundary condition. Consequently, the observables constructed by means of the equivariant coordinates \eqref{eq_coord0} and \eqref{eq:Ycoords} via \eqref{eq:ref3} are gauge-invariant with respect to diffeomorphisms which vanish at spatial infinity. As anticipated, the coordinates $Y_\Gamma^i$ are non-local, but the non-locality of $G_\phi $ is relatively harmless since its wave front set is that of the $\delta$-function, and renormalisation of expressions involving such objects is well under control, cf. Section \ref{sec:quantization}.

The coordinates \eqref{eq:Ycoords} are not entirely well-suited for practical computations because of the fact that the rescaled Mukhanov-Sasaki variable $\mathfrak{R}$ appears convoluted with the Coulomb potential. In order to remedy this we use a different family of spatial hypersurfaces and a corresponding modification of the spatial Laplacian and its inverse. To this end we consider a number of additional quantities related to the slicing induced by the time-function $\tilde \phi$: the lapse function $N_\phi$, the extrinsic curvature $K_{\phi,ab}$, and the spatial Ricci scalar $R^{(3)}_\phi $ which are defined and computed respectively as
\begin{gather}
\label{eq_lapsenongi}
N_\phi:= |\tilde{g}^{-1}_\lambda(d\tilde\phi,d\tilde \phi)|^{-1/2} = \frac{a}{\phi^\prime}\left(1 - \frac{\varphi'}{\phi'} + A\right)   + O(\delta\Gamma^2)\,,\\
K_{\phi,ab} := {h_{\phi,a}}^c\nabla_{c} n_{\phi,b}\,,\qquad K_\phi := {K_{\phi,a}}^a = \frac{3\HHH}{a} + O(\delta \Gamma)\,,\label{eq_traceK}\\
R^{(3)}_\phi  := K_{\phi,ab}K_\phi ^{ba}-K_\phi ^2+2\left(R_{ab}-\frac12 R \tilde g_{ab}\right)n_\phi^an_\phi^b  = \frac{4}{a^2} \Delta \RRR   + O(\delta\Gamma^2)\,, \label{eq_spatialCurv}
\end{gather}
where $n_\phi$ and $h_\phi$ are defined respectively in \eqref{eq_nphi} and \eqref{eq_hphi}. Using these quantities, we define a new time function
$$
\mathfrak{t}:=\tilde \phi - \frac{3 N_\phi}{4 K_\phi } G_\phi  R^{(3)}_\phi  = \phi + \frac{\phi^\prime}{\HHH} D+O(\delta\Gamma ^2)\,,
$$
If we define the spatial metric $h_\mathfrak{t}$, the Laplacian $\Delta_\mathfrak{t}$ and its inverse $G_\mathfrak{t}$ in analogy to $h_\phi$, $\Delta_\phi $ and $G_\phi $ by replacing $\tilde \phi$ with $\mathfrak{t}$ we obtain
\begin{equation}\label{eq:Xcoords}
X_\Gamma^i := \left(1-G_\mathfrak{t} \circ \Delta_\mathfrak{t}\right)x^i = x^i +  \partial_i  E+O(\delta \Gamma ^2)\,,
\end{equation}
and the spatial coordinates $X_\Gamma^i$ share the qualitative properties of the initially defined $Y_\Gamma^i$.

 
\section{Examples of gauge-invariant observables at second order}
\label{sec_obs}

In the previous sections we have developed a principle to construct gauge-invariant perturbative observables from non-gauge-invariant ones. In the following we demonstrate this principle at the example of two observables which are relevant in Cosmology. To this end we use the covariant coordinates \eqref{eq_coord0} and \eqref{eq:Xcoords}.

Despite the mild non-locality inherent in the covariant spatial coordinates $\eqref{eq:Xcoords}$, we are interested in observables $A_\Gamma$ which are local functionals of the configuration $\Gamma$. The non-locality of $\mathcal{A}_\Gamma = A_\Gamma \circ \alpha_\Gamma$ implied by the non-locality of $X^i_\Gamma$ in \eqref{eq:Xcoords} appears only because we consider the local functional $A_\Gamma$ relative to the non-local functional $X_\Gamma$. Since the background $\Gamma_0$ depends only on time the same applies to the background value of any local functional $A_\Gamma$. Consequently, at first order only the field $X^0_\Gamma$ \eqref{eq_coord0} chosen as time coordinate enters the formula for gauge-invariant fields. At second order also the fields used as spatial coordinates $X^i_\Gamma$ \eqref{eq:Xcoords} enter the expression. 

The inverse $J$ of the Jacobi matrix of the coordinate transform $X_{\Gamma_0}$ on the background is
\[
J=\left(\begin{array}{cccc}
\frac{1}{\phi'}&0&0&0\\
0&1&0&0\\
0&0&1&0\\
0&0&0&1
\end{array}\right).
\]
The field dependent shifts from Section \ref{sec:generalexpansion2} with respect to these coordinates up to second order are
\[
x_1^0=-\frac{\varphi}{\phi^{\prime}}\,,\qquad \ x_1^i=-\partial_i E\,,
\]
and
\[
x_2^{0}=-\frac{\phi''\ph^2}{(\phi')^3}+\frac{2}{\phi'}\left(\frac{\varphi'\varphi}{\phi'}+(\partial_i \varphi)\partial^i E\right),
\]
\[
x_2^{i}=\frac{2\varphi}{\phi'}\partial_i E^\prime +2(\partial^i \partial^j E)\partial_j E - (X^i_\Gamma - x^i - \partial_i E)\,.
\]

Thus, for a field $A_\Gamma$ whose value on the background depends only on time the contributions up to second order for the gauge-invariant modification $\mathcal{A}_\Gamma = A_\Gamma \circ \alpha_\Gamma$ are
\[
\mathcal{A}_0=A_0\,,\qquad \mathcal{A}_1=A_1-\frac{A_0'\varphi}{\phi'}\,,
\]
\[
\mathcal{A}_2=A_2-\frac{2A_1'\varphi}{\phi'}-2(\partial_iA_1)\partial^i E+A_0'\left(-\frac{\phi''\ph^2}{(\phi')^3}+\frac{2}{\phi'}\left(\frac{\varphi'\varphi}{\phi'}+(\partial_i \varphi)\partial^i E\right)\right)+\frac{A_0''\varphi^2}{(\phi')^2}\,.
\]

If we were to use the fields $Y^i_\Gamma$ \eqref{eq:Ycoords} as spatial coordinates rather than the fields $X^i_\Gamma$ \eqref{eq:Xcoords}, then the corresponding expression for $\mathcal{A}_1$ would remain unchanged whereas $\mathcal{A}_2$ would change by replacing all occurrences of $\partial_i E$ by $\partial_i E + G_\Delta \partial_i \mathfrak{R}$. This demonstrates the dependence of the gauge-invariant constructions on the chosen covariant coordinate system.

\subsection{The lapse function}
\label{sec_obs_lapse}

The Sachs-Wolfe effect is one of the main building blocks of the current understanding of the Cosmic Microwave Background (CMB). A rough estimate of this effect can be obtained using the Tolman idea, see e.g. \cite{Mukhanov:2003xr}. Given a spacetime with a (conformal) timelike Killing field $\kappa$ and a state in equilibrium relative to the $\kappa$-flow with absolute temperature $T$, an observer with four-velocity $u\propto \kappa$ measures the temperature $\widetilde T=T/N$ with $N$ denoting the lapse function $N=\sqrt{|g(\kappa,\kappa)|}$.

In the context of Cosmology we use the Klein-Gordon field $\tilde\phi$ as a time coordinate and consider the vector 
$$
\kappa_\phi := N_\phi n_\phi = \frac{1}{\phi'}\partial_\tau + O(\delta \Gamma)
$$
with $N_\phi$, $n_\phi$ defined in \eqref{eq_lapsenongi} and \eqref{eq_nphi} respectively as an approximate conformal Killing vector -- in the sense that $\mathcal{L}_{\kappa_\phi} \tilde g - 2 \HHH/\phi' \tilde g = O(\phi'', \delta \Gamma)$. The corresponding lapse function is $N_\phi = a/\phi^\prime + O(\delta \Gamma)$. Its background value is not vanishing and thus it is not automatically gauge-invariant at linear order.
 
As described in Section \ref{sec_covcoords}, we may obtain a non-perturbatively gauge-invariant version of the lapse function by setting and computing
\begin{align}
\NN_\phi:= & N_\phi \circ \alpha_\Gamma = \frac{a}{\phi^\prime}\left(1-\left((\partial_\tau + \mathcal{H})\frac{\varphi}{\phi^\prime} - A\right)\right)+O(\delta\Gamma^2)\label{eq_lapsegi}\\
=&\frac{a}{\phi^\prime}\left(1-\left((\partial_\tau + \mathcal{H})\frac{\chi}{\phi^\prime} - \Phi\right)\right)+O(\delta\Gamma^2)\,,\notag
\end{align}
where $\Phi$ and $\chi$ are the gauge-invariant fields reviewed in Section \ref{sec:perturbationsintro}. Using the on-shell identities \eqref{eq:onshellfirstorder}, \eqref{eq:onshellfirstorder2} and the definition of the Mukhanov-Sasaki field $\mu$ we can rewrite the linear term as
\begin{align*}
\NN_{\phi,1}=& \lambda\frac{a}{\phi^\prime}\left((\partial_\tau + \mathcal{H})\frac{\chi}{\phi^\prime} - \Phi\right)=\frac{a}{(\phi^\prime)^2}\left(\mu^\prime + \left(\frac{\mathcal{H}^\prime}{\mathcal{H}}-\frac{\phi^{\prime\prime}}{\phi^\prime}\right)\mu\right)\\
=&\frac{2a}{(\phi^\prime)^3} \Delta \Phi = -\frac{2a}{(\phi^\prime)^3} \Delta \Psi\,.
\end{align*}
Using the quantities introduces in Section \ref{sec:coordinates}, we may extract the Bardeen potential on-shell from $N_{\phi}$ as
$$
\left[\frac{1}{2 N_{\phi}^3} G^2_\phi \Delta_\phi N_{\phi}\right]\circ \alpha_\Gamma = \Phi + O(\delta \Gamma^2)\,.
$$
In fact, one could use the above equation as a covariant, gauge-invariant, all-order (and on shell) definition of $\Phi$; however, we shall refrain from doing so.

In order to display second order expressions in a readable form we omit terms containing the metric perturbation components $V_i$, $W_j$ and $T_{ij}$ and use once more the Bardeen potentials $\Phi$, $\Psi$ and the gauge-invariant scalar field perturbation $\chi$. We stress that the particular expressions of these fields at linear and higher order are not needed for the actual computations but just for a compact display of the result. Using this, we arrive at the following second order form of the gauge-invariant lapse function
\begin{align*}
\NN_{\phi,2} &= \frac{a}{\phi'}\left(-\Phi^2 -2 \left(\frac{\Phi \chi}{\phi'}\right)' -2 \mathcal{H}\frac{\Phi \chi}{\phi'}  +2 \left(\left(\frac{\chi}{\phi'}\right)'\right)^2+\left( \frac{\phi''}{\phi'}+2\mathcal{H} \right)\left(\frac{\chi^2}{\phi'^2}\right)' 
+
\right.
\\
&\qquad \left.+\left(\mathcal{H}^2+\mathcal{H}'+\frac{\phi'''}{\phi'}+\mathcal{H}\frac{\phi''}{\phi'}- \frac{\phi''^2}{\phi'^2}\right)\frac{\chi^2}{\phi'^2} + \sum^3_{i=1}\left(\partial_i\left(\frac{\chi}{\phi'}\right)\right)^2
+2\frac{\chi}{\phi'}\left(\frac{\chi}{\phi'}\right)''
\right),
\end{align*}
where, as before, we use the notation that e.g. $\Phi = \lambda \Phi_1 + \frac12 \lambda^2 \Phi_2+O(\lambda^3)$ and omit the second order terms linear in $\Phi$, $\chi$ displayed already in \eqref{eq_lapsegi}.

\subsection{The spatial curvature}  A further observable of interest is the scalar curvature of the spatial metric induced by a particular slicing because for a large class of slicings this quantity vanishes in the background and thus is automatically gauge-invariant at linear order. Moreover, for the slicing defined by the inflation field it is related to the Mukhanov-Sasaki field $\mu$ which has a very simple dynamical equation.

We have already discussed the spatial curvature relative to the slicing induced by $\tilde \phi$. It may be computed as \eqref{eq_spatialCurv}
$$
R^{(3)}_\phi  =  \frac{4}{a^2} \Delta \RRR   + O(\delta\Gamma^2)\,,\qquad \mathfrak{R} = \frac{\HHH}{\phi^\prime}\mu =\frac{\HHH}{\phi^\prime}\varphi - D \,.
 $$
In the literature, the quantity $\RRR$ is usually called the \emph{comoving curvature perturbation}. This is due to the fact that the $\tilde \phi$-slicing may be equivalently characterised by the condition that 
$$
T(\tilde \phi)_{ab} n_\phi^a = -\tilde g_{ab} n_\phi^a T(\tilde \phi)_{cd} n_\phi^c n_\phi^d\,,
$$
i.e. that the energy flux of $\tilde \phi$ is parallel to $n_\phi$, where  $T(\tilde \phi)_{ab}$ is the stress tensor of $\tilde \phi$. 

An alternative slicing considered in the literature is the one defined by the energy density $\tilde\rho$ of $\tilde \phi$
$$
\tilde\rho := T(\tilde \phi)_{ab} n_\phi^a n_\phi^b = \rho + \varrho\,,
$$
$$
\rho := \frac{(\phi')^2}{2 a^2}\,,\qquad \varrho := V^{(1)}(\phi)\varphi + \frac{\phi'(\varphi' - \phi' A)}{a^2}+O(\delta \Gamma^2)\,.$$
The spatial curvature $R^{(3)}_\rho$ with respect to this slicing, defined in analogy to $R^{(3)}_\phi$, reads
$$
R^{(3)}_\rho = \frac{4}{a^2}\Delta \zeta + O(\delta \Gamma)\,,\qquad \zeta := \frac{\HHH}{\rho'}\varrho-D \,,
$$
where $\zeta$ is called \emph{uniform density perturbation} because $\tilde \rho$ is by definition constant on the hypersurfaces in the slicing relative to $\tilde \rho$. The global sign in the definition of $\zeta$ is conventional.

As anticipated, the background contributions of $R^{(3)}_\phi$ and $R^{(3)}_\rho$ vanish and thus 
$$
\R^{(3)}_\phi := R^{(3)}_\phi \circ \alpha_\Gamma = R^{(3)}_\phi + O(\delta\Gamma^2)\,,\qquad \R^{(3)}_\rho := R^{(3)}_\rho \circ \alpha_\Gamma = R^{(3)}_\rho + O(\delta\Gamma^2)\,,
$$
cf. \eqref{eq_gaugeinvexp1}, \eqref{eq_gaugeinvexp2}. In order to display the second order contribution to $\R^{(3)}_\phi$, we make the simplifications discussed for the lapse function in Section \ref{sec_obs_lapse}. Proceeding like this, we find
\begin{align}\label{eq_spatialgi}
\R^{(3)}_{\phi,2}=&\frac{8}{a^2}\left(\Delta\left(2\RRR^2 -  \frac{\chi}{\phi'}(\partial_\tau + 2\HHH)\RRR + \frac12\left(\HHH'+2\HHH^2-\frac{\HHH \phi''}{\phi'}\right)\left(\frac{\chi}{\phi'}\right)^2\right)\right.\\
&\qquad\qquad \left.- \frac{5(\partial_i \RRR) \partial^i \RRR}{2}\right).\notag
\end{align}

We omit the result for $\R^{(3)}_{\rho,2}$ computed with the coordinate system $X_\Gamma$ defined in \eqref{eq_coord0} and \eqref{eq:Xcoords}, because it is rather long due to the ``mismatch'' between the time coordinate $\tilde \phi$ used in $X^0_\Gamma$ and the time coordinate $\tilde \rho$ used in the definition of $R^{(3)}_\rho$. Clearly, using $\tilde \rho$ as a time coordinate in both aspects we would obtain a second order expression $\R^{(3)}_{\rho,2}$ which is of the form \eqref{eq_spatialgi} up to the replacements
\begin{equation}\label{eq_replacements}
\RRR \mapsto \zeta\,,\qquad \phi\mapsto \rho\,,\qquad \chi\mapsto \pi:= V^{(1)}(\phi)\chi + \frac{\phi'(\chi'-\phi'\Phi)}{a^2}\,,
\end{equation}
where $\pi$ is gauge-invariant with $\pi = \varrho + O(\delta \Gamma^2)$ in the longitudinal gauge.

On shell and at first order, $\mu$, and thus $\RRR$, are preferred observables because they have canonical equal-time Poisson brackets and thus in the quantized theory they commute at spacelike separations, in contrast to $\Psi$, $\Phi$ and $\chi$ \cite{Eltzner, Hack}. Moreover, again on shell and at first order, one may compute
$$
\zeta = \RRR -\frac{2 \Delta \Phi}{3 (\phi')^2} = \RRR - \frac{\RRR'}{3 \HHH}\,.
$$
Consequently, $\zeta$ shares the causality properties of $\mu$ and $\RRR$.

Apart from the phenomenological relevance of an all-order definition of $\RRR$, $\mu$ and $\zeta$, it is interesting on conceptual grounds to investigate whether the causality property of these fields persists at higher orders. To this end, we need a fully covariant and gauge-invariant all-order definition of $\RRR$, $\mu$ and $\zeta$. Such a definition may be given by means of covariant quantities introduced in Section \ref{sec:coordinates}:
\begin{align}
\label{eq_higherorderR}
\left[\frac{1}{4} G_\phi R^{(3)}_\phi\right]\circ \alpha_\Gamma &= \frac{a^2}{4} G_\Delta \mathcal{R}^{(3)}_\phi- a^2 G_\Delta  \delta\Delta \RRR +  O(\delta\Gamma^3)\notag\\
&=\HHH\frac{\chi}{\phi'}-\Psi + \RRR^2 - 2 \HHH\frac{\chi}{\phi'}\RRR + \frac12\left(\HHH'+2\HHH^2-\frac{\HHH \phi''}{\phi'}\right)\left(\frac{\chi}{\phi'}\right)^2 + \\
\notag&\qquad + G_\Delta\left(\frac{(\partial_i \RRR) \partial^i \RRR}{2}\right) +  O(\delta\Gamma^3)\,,
\end{align}
\begin{equation}
\label{eq_higherordermu}
\left[\frac{3 N_\phi}{4 K_\phi } G_\phi  R^{(3)}_\phi\right]\circ \alpha_\Gamma = \mu + O(\delta\Gamma^2)\,,\qquad \left[\frac{1}{4} G_\phi R^{(3)}_\rho\right]\circ \alpha_\Gamma = \zeta + O(\delta\Gamma^2)\,.
\end{equation}

In \eqref{eq_higherorderR} we wrote the $O(\delta \Gamma)$ term as $\HHH \chi/\phi' - \Psi$ instead of $\RRR$ because the fields $\chi$, $\Psi$ are defined in such a way that they are invariant also with respect to second order gauge transformations (cf. the end of Section \ref{sec:perturbationsintro}), whereas $\RRR=\HHH \varphi/\phi' - D$ is only gauge-invariant up to the first order. 

In analogy to our discussion of $R^{(3)}_\rho$, using $\tilde \rho$ rather than $\tilde \phi$ both as the time coordinate $X^0_\Gamma$ and as the time function defining a foliation of spacetime, we obtain a higher order definition of $\zeta$ which is of the form \eqref{eq_higherorderR} up to the replacements in \eqref{eq_replacements} (whereby a second order generalisation of $\pi$, which can be constructed in analogy to the second order Bardeen potentials, is needed).

In the literature, several possible second order gauge-invariant corrections to $\RRR$ are considered. One often encounters constructions where in a gauge with $\varphi=0$ (or $D=0$), the second order corrections to $\RRR$ vanish -- at least in situations where spatial derivatives can be neglected in comparison to temporal ones, see e.g. \cite{Maldacena:2002vr, Malik:2008im, Prokopec:2012ug, Vernizzi:2004nc}. In fact $\RRR$ is often defined by the condition $\RRR = -D$ in a gauge where $\varphi=0$. A quick analysis reveals that this is not the case in our construction \eqref{eq_higherorderR}. In \cite{Vernizzi:2004nc} it is argued that expressions for $\RRR$ valid up to second order that are not of this form, e.g. the one in \cite{Acquaviva:2002ud}, are potentially physically ill-behaved because they are not conserved on ``super-Hubble scales''. Here, conservation of a function $f(\tau,\vec x)$ on ``super-Hubble scales'' means that the Fourier transform $\hat f(\tau,\vec k)$ of $f$ with respect to $\vec x$ satisfies $\partial_\tau \hat f(\tau,\vec k) = O(|\vec k|/\HHH)$. This property, whose relevance is explained e.g. in \cite{Maldacena:2002vr, Vernizzi:2004nc}, usually holds only on-shell. It would be interesting to check whether our result for $\RRR$ as given in \eqref{eq_higherorderR} (and the analogous result for $\zeta$) is conserved in this sense; however, this is beyond the scope of the present work. 

\section{Quantization}
\label{sec:quantization}

In the previous sections we have prepared the ground for an all-order perturbative quantization of the Einstein-Klein-Gordon system on FLRW backgrounds, i.e. for a conceptually clear higher-order generalisation of quantized cosmological perturbation theory. In this section we would like to sketch the steps necessary for a full construction of the quantum theory. A detailed account will be given in a future work \cite{longpaper}.

\subsection{BRST quantization}

It is known that a direct quantization of non-linear gauge-invariant observables in a theory with local gauge symmetries is difficult. The standard way out is to perform a gauge fixing in the sense of the BRST method, or more generally, the BV formalism, as treated in \cite{Hollands, Fredenhagen:2011an, FR-BVQ}. There one adds a Fermionic vector field $c^{\mu}$ (the ghost field), which describes the infinitesimal gauge transformations, auxiliary scalar fields $b_{\mu}, \bar{c}_{\mu}$, where $b_{\mu}$ (the Nakanishi-Lautrup field) is Bosonic and  $\bar{c}_{\mu}$ (antighost) is Fermionic, $\mu=0,\ldots,3$. Infinitesimal coordinate transformations are described by the BRST operator $s$, which acts on scalar local functionals $A$ of the metric, the inflaton and the $b$ fields  
by
\[s(A)(x)=c^{\mu}(x)\partial_{\mu}A(x)\ ,\]
on the components of the ghost field by
\[s(c^{\mu})(x)=c^{\nu}(x)\partial_{\nu}c^{\mu}(x)\ ,\]
on antighosts by
\[s(\bar{c}_{\mu})(x)=ib_{\mu}(x)-c^{\nu}(x)\partial_{\nu}\bar{c}_{\mu}(x)\]
and satisfies on products the graded Leibniz rule so that $s^2=0$. One can characterise the classical observables as functionals in the kernel of $s$ modulo those in the image of $s$ (i.e. classical observables belong to the $0$-th cohomology group of $s$). 

The field equations for the extended system are the usual field equation for $\tilde\phi$ as well as
\[R_{\mu\nu}=T(\tilde\phi)_{\mu\nu}-\frac12 T(\tilde\phi)\tilde g_{\mu\nu}+s(i\partial_{(\mu}\bar{c}_{\nu)})\]
\[\square_{\tilde g} c^{\mu}=0\]
\[\square_{\tilde g} \bar{c}_{\mu}=0\]
\[|\mathrm{det}\tilde g|^{-\frac12}\partial_{\mu}|\mathrm{det}\tilde g|^{\frac12}\tilde g^{\mu\nu}=\kappa^{\mu\nu}b_{\mu}\,.\]
Here $\kappa$ is a non-degenerate fixed tensor.

The quantization of the extended system now proceeds largely analogous to the pure gravity treatment in \cite{BFR}. The main idea is to use deformation quantization to deform the algebra of functionals as well as the BRST operator $s$. Elements of the cohomology of the quantized (i.e. deformed) BRST operator $s$ are then interpreted as quantized versions of the gauge-invariant fields discussed in the previous sections. 

\subsection{Renormalisation}

A conceptual and technical difference to the pure gravity case treated in \cite{BFR} arises because of the fact that we have introduced a mild non-locality via the non-local spatial coordinates $X^i_\Gamma$ \eqref{eq:Xcoords}. In \cite{BFR} renormalisation was treated in the Epstein-Glaser framework which is initially only suitable for local functionals. As we have to deal with non-local expressions, we need to extend this framework from local quantities to non-local ones. Recall that
\[
X_{\Gamma}^i=(1-G_\phi \Delta_\phi )x^i = \sum_{k=0}^\infty (-G_0\,\delta \Delta)^k x^i\,,
\]
where $\Delta_\phi =\Delta_0+\delta \Delta $ is the Laplacian relative to the $\tilde \phi$-slicing and $G_\phi =\sum_{k=0}^\infty (-G_0\,\delta \Delta)^kG_0$ is its Green's function for suitable boundary conditions, cf. Section \ref{sec:coordinates}. 

Our gauge-invariant observables can be expanded as Taylor series in $X_\Gamma^a$, so in order to discuss the renormalisation of non-local contributions it is sufficient to discuss the kind of singularities that arise from considering the time-ordered products involving $X_\Gamma^i$. The general strategy is similar to the standard setting. We start with non-renormalised expressions where the $n$-fold time-ordered product involving $X_\Gamma^i$ and local functionals $F_1$,\dots,  $F_{n-1}$ is given by
\[
\mathcal{T}_n(X_\Gamma^i,F_1,\dots,F_{n-1}):= m\circ e^{\hbar\sum_{0\leq k<l\leq n-1}D_{\mathrm{F}}^{kl}}(X_\Gamma^i\otimes F_1\otimes\dots\otimes F_{n-1})\,.
\]
where $m$ denotes pointwise multiplication and $D_{\mathrm{F}}^{kl}\doteq \langle\Delta_{S_0}^{\mathrm{F}},\frac{\delta^2}{\delta\Gamma_k\delta\Gamma_l}\rangle$ with $\Delta_{S_0}^{\mathrm{F}}$ denoting the Feynman propagator of the full linearised theory. For simplicity, we suppress all indices. This expression is then expanded into graphs. The non-locality is expressed by the fact that our graphs have now two kinds of vertices and two kinds of propagators. Namely, there are the ``usual'' Feynman propagators of the theory (for simplicity all denoted by 
\begin{tikzpicture}
\useasboundingbox  (-0.07,0) rectangle +(1.2,-0.02);
\draw [line width=0.30mm] (0,0.1) -- (1,0.1);
\end{tikzpicture}
), but also the ``internal'' propagators $G_0$ corresponding to lines \begin{tikzpicture}
\useasboundingbox  (-0.07,0) rectangle +(1.2,-0.05);
\draw [line width=0.30mm] (0,0.13) -- (1,0.13);
\draw [line width=0.30mm] (0,0.08) -- (1,0.08);
\end{tikzpicture}. 

As for the vertices, there are the external vertices \begin{tikzpicture}
\useasboundingbox  (-0.07,0) rectangle +(0.1,-0.1);
\filldraw circle (1.5pt);
\end{tikzpicture} arising from local functionals $F_{1},\dots,F_{n-1}$ and from the vertex corresponding to the explicit spacetime dependence of $X^i_\Gamma$, but also the internal vertices \begin{tikzpicture}
\useasboundingbox  (-0.07,0) rectangle +(0.1,-0.1);
\draw circle (1.5pt);
\end{tikzpicture} obtained from the $\delta \Delta$ operators. An example contribution would be
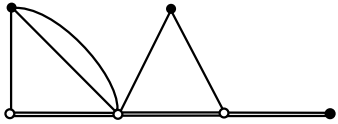
\begin{center}
	\begin{tikzpicture}[scale=0.7,x=1.00mm, y=1.00mm, inner xsep=0pt, inner ysep=0pt, outer xsep=0pt, outer ysep=0pt]
	\path[line width=0mm] (36.57,47.12) rectangle +(66.49,25.85);
	\definecolor{L}{rgb}{0,0,0}
	\definecolor{F}{rgb}{0,0,0}
	\path[line width=0.30mm, draw=L] (39.40,50.07) circle (0.84mm);
	\path[line width=0.30mm, draw=L] (59.92,49.95) circle (0.84mm);
	\path[line width=0.30mm, draw=L] (80.07,50.20) circle (0.84mm);
	\path[line width=0.30mm, draw=L, fill=F] (100.22,50.07) circle (0.84mm);
	\path[line width=0.30mm, draw=L] (40.24,50.28) -- (59.14,50.23);
	\path[line width=0.30mm, draw=L] (39.94,49.62) -- (59.18,49.63);
	\path[line width=0.30mm, draw=L] (60.77,50.31) -- (79.30,50.31);
	\path[line width=0.30mm, draw=L] (60.87,49.69) -- (79.36,49.71);
	\path[line width=0.30mm, draw=L] (80.88,50.28) -- (99.45,50.24);
	\path[line width=0.30mm, draw=L] (80.58,49.62) -- (99.52,49.64);
	\path[line width=0.30mm, draw=L, fill=F] (39.73,70.23) circle (0.74mm);
	\path[line width=0.30mm, draw=L] (39.63,69.62) -- (39.63,50.74);
	\path[line width=0.30mm, draw=L] (39.83,70.23) -- (59.33,50.64);
	\path[line width=0.30mm, draw=L] (39.83,70.15) .. controls (47.90,71.01) and (60.00,57.62) .. (59.83,50.93) .. controls (59.83,50.89) and (59.37,50.64) .. (59.40,50.67);
	\path[line width=0.30mm, draw=L, fill=F] (70.01,70.00) circle (0.74mm);
	\path[line width=0.30mm, draw=L] (70.01,69.78) -- (60.44,50.58);
	\path[line width=0.30mm, draw=L] (69.90,69.89) -- (79.76,51.04);
	\end{tikzpicture}%
\end{center}

To see that such graphs can be renormalised, consider the simplest divergent case, namely
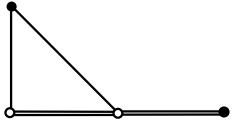
\begin{center}
	\begin{tikzpicture}[scale=0.7,x=1.00mm, y=1.00mm, inner xsep=0pt, inner ysep=0pt, outer xsep=0pt, outer ysep=0pt]
	\path[line width=0mm] (36.57,47.12) rectangle +(66.49,25.85);
	\definecolor{L}{rgb}{0,0,0}
	\definecolor{F}{rgb}{0,0,0}
	\path[line width=0.30mm, draw=L] (39.40,50.07) circle (0.84mm);
	\path[line width=0.30mm, draw=L] (59.92,49.95) circle (0.84mm);
	\path[line width=0.30mm, draw=L, fill=F] (80.07,50.20) circle (0.84mm);
	\path[line width=0.30mm, draw=L] (40.24,50.28) -- (59.14,50.23);
	\path[line width=0.30mm, draw=L] (39.94,49.62) -- (59.18,49.63);
	\path[line width=0.30mm, draw=L] (60.77,50.31) -- (79.30,50.31);
	\path[line width=0.30mm, draw=L] (60.87,49.69) -- (79.36,49.71);
	\path[line width=0.30mm, draw=L, fill=F] (39.73,70.23) circle (0.74mm);
	\path[line width=0.30mm, draw=L] (39.63,69.62) -- (39.63,50.74);
	\path[line width=0.30mm, draw=L] (39.83,70.23) -- (59.33,50.64);
	\end{tikzpicture}%
\end{center}
The kernel of $G_0$ considered as a distribution on $M^2$ is of the form
$$
G_0(x,y) = c(\tau_x) \delta(\tau_x,\tau_y) \frac{1}{|\vec {x}-\vec{y}|}
$$
with a smooth function $c$. The wave front set of $G_0(x,y)$ is the one of $\delta(x,y)$ and its scaling degree is 2. The vertex operators $\delta \Delta$ are differential operators of at most second order. By direct inspection we thus see that the only singularity of the loop in the above example is at the total diagonal 
and by power counting we find that the degree of divergence of this loop is at most 2, so that the appropriately renormalised expression is unique up to at most two derivatives of $\delta$ distributions of the three loop vertices. In general the degree of divergence of a loop containing ``internal'' propagators may be higher or lower than in the above example depending on the number of Feynman propagators appearing in the loop; the same applies to the renormalisation freedom of general loops.

These arguments indicate that the new types of graphs do not create new problems in the UV regime. We briefly sketch why we do not expect additional IR problems. We have already pointed out that our setup is only meaningful if we restrict the admissible classical configurations to those which vanish at spatial infinity. By consistency we need the same behaviour for the correlation functions of the quantized theory, in particular for the Feynman propagators of the linearised model. Provided quantum states (or more general Hadamard parametrices) with this property  exist -- this is not obvious and needs to be proven -- we expect that the integrals corresponding to the ``internal'' vertices will converge.

The remaining problem is to deal with the combinatorics of such graphs and ensure that the renormalisation can be performed systematically order by order. This can be done by a slight generalisation of the standard framework and will be discussed in detail in our forthcoming paper \cite{longpaper}. In the same publication we will also prove the validity of Ward identities analogous to the ones proven by Hollands for the Yang-Mills theory \cite{Hollands}.

\section{Conclusions} 
We described how cosmological perturbation theory may be derived from a full theory of perturbative quantum gravity.
This demonstrates that perturbative quantum gravity can already be tested by present observations. Moreover, on a more practical side, our definition of gauge-invariant observables provides a conceptually simple way of extending the observables which are relevant for the interpretation of cosmological observations to arbitrary high orders. 

However, even in linear order, our discussion clarifies the choice of good observables, as we have indicated at the example of the lapse function $N_\phi$ with respect to the spatial hypersurfaces of constant inflaton field. Initially $N_\phi$ is not gauge-invariant, but our construction yields a gauge-invariant version which at linear order and on shell may be expressed in terms of the Bardeen potential $\Phi$ that is related to the temperature fluctuations of the CMB via the Sachs-Wolfe effect.

We computed examples of gauge-invariant observables beyond linear order and found a second-order expression for the comoving curvature perturbation which seems to differ from constructions in other works. As in the literature there is some debate about whether some constructions are physically well-behaved, see. e.g. \cite{Vernizzi:2004nc}, it would be interesting to investigate the physical properties of our result, even though it is clear from the outset that it has a transparent geometric interpretation.

Finally we have sketched the details of the quantization of the Einstein-Klein-Gordon system on cosmological backgrounds beyond linear order. We believe that the strategy outlined here leads to a full renormalised all-order theory of cosmological perturbations by means of which higher order corrections to standard results in cosmology may be computed.


\vspace{1cm}
\noindent{\bfseries Acknowledgements}\\

K.F., N.P. and K.R. would like to thank the Erwin Schr\"odinger Institute in Vienna, where part of the research reported here was carried out, for the kind hospitality.



\begin{thebibliography}{99}

\bibitem[ABMR03]{Acquaviva:2002ud}
  V.~Acquaviva, N.~Bartolo, S.~Matarrese and A.~Riotto: Second order cosmological perturbations from inflation,
  Nucl.\ Phys.\ B {\bf 667} (2003) 119
  doi:10.1016/S0550-3213(03)00550-9
  [astro-ph/0209156].
  
\bibitem[BFR15]{BFR}R. Brunetti, K. Fredenhagen, K. Rejzner: Quantum gravity from the point of view of
locally covariant quantum field theory, to appear in Commun. Math. Phys., arXiv:1306.1058 [math-ph].

\bibitem[BFHPR]{longpaper} R. Brunetti, K. Fredenhagen, T.-P. Hack, N. Pinamonti, K. Rejzner, in preparation.

\bibitem[BMR07]{Bartolo:2006fj}
  N.~Bartolo, S.~Matarrese and A.~Riotto: CMB Anisotropies at Second-Order. 2. Analytical Approach,
  JCAP {\bf 0701} (2007) 019
  doi:10.1088/1475-7516/2007/01/019
  [astro-ph/0610110].

\bibitem[BK95]{Brown:1994py}
  J.~D.~Brown and K.~V.~Kuchar: Dust as a standard of space and time in canonical quantum gravity,
  Phys.\ Rev.\ D {\bf 51} (1995) 5600
  doi:10.1103/PhysRevD.51.5600
  [gr-qc/9409001].

\bibitem[BMMS97]{Bruni:1996im}
  M.~Bruni, S.~Matarrese, S.~Mollerach and S.~Sonego: Perturbations of space-time: Gauge transformations and gauge invariance at second order and beyond,
  Class.\ Quant.\ Grav.\  {\bf 14} (1997) 2585
  doi:10.1088/0264-9381/14/9/014
  [gr-qc/9609040].
 
\bibitem[Di05]{Dittrich:2005kc}
  B.~Dittrich: Partial and complete observables for canonical general relativity,
  Class.\ Quant.\ Grav.\  {\bf 23} (2006) 6155
  doi:10.1088/0264-9381/23/22/006
  [gr-qc/0507106]. 
 
\bibitem[DF01]{DF} M. D\"utsch, K. Fredenhagen: Perturbative algebraic field theory, and deformation quantization,
Proceedings of the Conference on Mathematical Physics in Mathematics and
Physics, Siena June 20-25 2000, [arXiv:hep-th/0101079].
 
\bibitem[El13]{Eltzner}B.~Eltzner: Quantization of Perturbations in Inflation, arXiv:1302.5358 [gr-qc].

\bibitem[FR12]{Fredenhagen:2011an}
  K.~Fredenhagen and K.~Rejzner: Batalin-Vilkovisky formalism in the functional approach to classical field theory,
  Commun.\ Math.\ Phys.\  {\bf 314} (2012) 93
  doi:10.1007/s00220-012-1487-y
  [arXiv:1101.5112 [math-ph]].

\bibitem[FR13]{FR-BVQ}K.~Fredenhagen and K.~Rejzner: Batalin-Vilkovisky formalism in perturbative algebraic quantum field theory,
  Commun.\ Math.\ Phys.\  {\bf 317} (2013) 697
  doi:10.1007/s00220-012-1601-1
  [arXiv:1110.5232 [math-ph]]. 
 
\bibitem[FR15]{FR-Advances in AQFT}K. Fredenhagen, R. Rejzner: Perturbative Construction of Models of Algebraic Quantum Field Theory, in: R.~Brunetti, C.~Dappiaggi, K.~Fredenhagen and J.~Yngvason (eds.): Advances in algebraic quantum field theory, Springer (2015), doi:10.1007/978-3-319-21353-8, arXiv:1503.07814 [math-ph].


\bibitem[Ha14]{Hack}T.~P.~Hack: Quantization of the linearized Einstein-Klein-Gordon system on arbitrary backgrounds and the special case of perturbations in inflation,
  Class.\ Quant.\ Grav.\  {\bf 31} (2014) no.21,  215004
  doi:10.1088/0264-9381/31/21/215004
  [arXiv:1403.3957 [gr-qc]].

\bibitem[Ho08]{Hollands} S.~Hollands: Renormalized Quantum Yang-Mills Fields in Curved Spacetime,
  Rev.\ Math.\ Phys.\  {\bf 20} (2008) 1033
  doi:10.1142/S0129055X08003420
  [arXiv:0705.3340 [gr-qc]].

\bibitem[Kh15]{Igor}I.~Khavkine: Local and gauge-invariant observables in gravity,
  Class.\ Quant.\ Grav.\  {\bf 32} (2015) no.18,  185019
  doi:10.1088/0264-9381/32/18/185019
  [arXiv:1503.03754 [gr-qc]].

\bibitem[LV10]{Langlois:2010vx}
  D.~Langlois and F.~Vernizzi: A geometrical approach to nonlinear perturbations in relativistic cosmology,
  Class.\ Quant.\ Grav.\  {\bf 27} (2010) 124007
  doi:10.1088/0264-9381/27/12/124007
  [arXiv:1003.3270 [astro-ph.CO]].
  
\bibitem[Ma03]{Maldacena:2002vr}
  J.~M.~Maldacena: Non-Gaussian features of primordial fluctuations in single field inflationary models,
  JHEP {\bf 0305} (2003) 013
  doi:10.1088/1126-6708/2003/05/013
  [astro-ph/0210603].  

\bibitem[MW09]{Malik:2008im}
  K.~A.~Malik and D.~Wands: Cosmological perturbations,
  Phys.\ Rept.\  {\bf 475} (2009) 1
  doi:10.1016/j.physrep.2009.03.001
  [arXiv:0809.4944 [astro-ph]].
  
\bibitem[Mu04]{Mukhanov:2003xr}
  V.~F.~Mukhanov: CMB-slow, or how to estimate cosmological parameters by hand,
  Int.\ J.\ Theor.\ Phys.\  {\bf 43} (2004) 623
  doi:10.1023/B:IJTP.0000048168.90282.db
  [astro-ph/0303072].

\bibitem[Na07]{Nakamura:2004rm}
  K.~Nakamura: Second-order gauge-invariant cosmological perturbation theory: Einstein equations in terms of gauge-invariant variables,
  Prog.\ Theor.\ Phys.\  {\bf 117} (2007) 17
  doi:10.1143/PTP.117.17
  [gr-qc/0605108].
  
\bibitem[Na14]{Nakamura:2014kza}
  K.~Nakamura: Recursive structure in the definitions of gauge-invariant variables for any order perturbations,
  Class.\ Quant.\ Grav.\  {\bf 31} (2014) 135013
  doi:10.1088/0264-9381/31/13/135013
  [arXiv:1403.1004 [gr-qc]].  

\bibitem[NH04]{Noh:2004bc}
  H.~Noh and J.~c.~Hwang: Second-order perturbations of the Friedmann world model,
  Phys.\ Rev.\ D {\bf 69} (2004) 104011.
  doi:10.1103/PhysRevD.69.104011

\bibitem[NH13]{Hwang:2012aa}
  J.~c.~Hwang and H.~Noh: Fully nonlinear and exact perturbations of the Friedmann world model,
  Mon.\ Not.\ Roy.\ Astron.\ Soc.\  {\bf 433} (2013) 3472
  doi:10.1093/mnras/stt978
  [arXiv:1207.0264 [astro-ph.CO]].

\bibitem[PW12]{Prokopec:2012ug}
  T.~Prokopec and J.~Weenink: Uniqueness of the gauge invariant action for cosmological perturbations,
  JCAP {\bf 1212} (2012) 031
  doi:10.1088/1475-7516/2012/12/031
  [arXiv:1209.1701 [gr-qc]].

\bibitem[Re98]{Reuter}M.~Reuter: Nonperturbative evolution equation for quantum gravity,
  Phys.\ Rev.\ D {\bf 57} (1998) 971
  doi:10.1103/PhysRevD.57.971
  [hep-th/9605030].
  
\bibitem[RS02]{Reuter2}M.~Reuter and F.~Saueressig: Renormalization group flow of quantum gravity in the Einstein-Hilbert truncation,
  Phys.\ Rev.\ D {\bf 65} (2002) 065016
  doi:10.1103/PhysRevD.65.065016
  [hep-th/0110054].
  
\bibitem[Ro02]{Rovelli:2001bz}
  C.~Rovelli: Partial observables,
  Phys.\ Rev.\ D {\bf 65} (2002) 124013
  doi:10.1103/PhysRevD.65.124013
  [gr-qc/0110035].  
  
\bibitem[SB98]{Sonego:1997np}
  S.~Sonego and M.~Bruni: Gauge dependence in the theory of nonlinear space-time perturbations,
  Commun.\ Math.\ Phys.\  {\bf 193} (1998) 209
  doi:10.1007/s002200050325
  [gr-qc/9708068]. 
  
\bibitem[Th06]{Thiemann:2004wk}
  T.~Thiemann: Reduced phase space quantization and Dirac observables,
  Class.\ Quant.\ Grav.\  {\bf 23} (2006) 1163
  doi:10.1088/0264-9381/23/4/006
  [gr-qc/0411031].  

\bibitem[Ve04]{Vernizzi:2004nc}
  F.~Vernizzi: On the conservation of second-order cosmological perturbations in a scalar field dominated Universe,
  Phys.\ Rev.\ D {\bf 71} (2005) 061301
  doi:10.1103/PhysRevD.71.061301
  [astro-ph/0411463].

\end{thebibliography}
\end{document}